# Content-Type Coding

Linqi Song  
UCLA  
Email: songlinqi@ucla.edu

Christina Fragouli  
UCLA and EPFL  
Email: christina.fragouli@ucla.edu

*Abstract*—This paper is motivated by the observation that, in many cases, we do not need to serve specific messages, but rather, any message within a content-type. Content-type traffic pervades a host of applications today, ranging from search engines and recommender networks to newsfeeds and advertisement networks. The paper asks a novel question: if there are benefits in designing network and channel codes specifically tailored to content-type requests. It provides three examples of content-type formulations to argue that, indeed in some cases we can have significant such benefits.

## I. INTRODUCTION

Coding traditionally aims to securely and efficiently convey *specific* information messages to one or more receivers. This broad aim encompasses most of the work in the field, from the channel coding theorem of Shannon [1], to recent breakthroughs in channel coding [2], [3], network coding [4], and index coding [5]. However, communication networks today are increasingly used to serve a fundamentally different traffic, that delivers *type of content* rather than specific messages. As a simple example, when we use the Internet to access our bank account, we ask and want to see very specific information. However, if we search for a photo of a humming bird, we do not care which specific humming bird photos we receive - we do not even know what humming bird pictures are available - we only care about the content type, that it is a humming bird and not an owl photo.

Content-type traffic pervades a host of applications today. For instance, content-delivery networks, such as the Akamai network, in many cases do not need to satisfy message-specific requests, but instead content-type requests (eg., latest news on Greece, popular sites on CNN.com, etc); smart search engines and recommendation systems (eg. Google, Pandora) generate in the majority content-type traffic; advertisement networks (eg. serving adds on hotels or cars), and newsfeeds on social networks (eg., cultural trends, following celebrities) also fall in the content-type category. The fact that content forms a significant percentage of the Internet traffic has been well recognized especially in the networking community; however, most of the work looks at what to replicate, where and how to store and from where to retrieve specific data.

We ask a very different question: are there benefits in designing network and channel codes specifically tailored to content-type traffic? We make the case that, indeed, if we realize that we need to serve content-type rather than specific messages, we can in some cases achieve significant benefits. The fundamental reason content type coding helps, is that it offers an additional dimension of freedom: we can select which, within the content type, specific message to serve, to optimize different criteria. For instance, we can create more coding opportunities, transform multiple unicast to a multicast session, and adapt to random erasure patterns, all of which can lead to higher throughput.

For different content-type coding problems, there are different mathematical representation. It is non-trivial to find a consistent mathematical model for a general content-type coding problem and it is beyond the scope of this paper. In this paper, we study the content-type coding problems by providing several examples to show the significant benefits of content-type coding over conventional message-specific coding. First, we provide an example of benefits over networks: we consider a classical example in the network coding literature, the combination network, and show the benefits in a content-type setup. Second, we provide an example of benefits over lossy networks: we consider a broadcast erasure channel with feedback setting, where a source wants to send messages to two receivers. Third, we review an example within the index coding setup, termed pliable coding, that we presented in a previous work, introduce an algebraic framework for pliable coding and use this framework to provide a novel lower bound.

The paper is organized as follows. Section II considers content-type coding over the combination network; Section III looks at broadcast erasure networks; Section IV introduces an algebraic framework for pliable index coding; and Section V concludes the paper with a short discussion.

## II. CONTENT-TYPE CODING OVER NETWORKS

*Motivating example:* Fig. 1 provides a simple example of content-type coding benefits. We start from the classical butterfly network, with two sources, $s_1$ and $s_2$, two receivers, $r_1$ and $r_2$, and unit capacity directed links, where each source has 2 messages of a given type. In the example, source $s_1$ (say advertising hotels) has 2 unit rate messages, $\{b_{11}, b_{12}\}$, and source $s_2$ (say advertising car rentals) also has 2 unit rate messages $\{b_{21}, b_{22}\}$. The min-cut to each receiver is two, thus she used all the network resources by herself, we could send to her two specific messages, for instance, to $r_1$: $b_{11}$ and $b_{21}$ and to $r_2$: $b_{12}$ and $b_{22}$. However, if we want to send to both receivers these specific requests, it is clearly not possible, as there are no coding opportunities; we can satisfy one receiver but not both. In contrast, if each receiver requires one message from each type, then we can multicast say $b_{11}$ and $b_{21}$ to both receivers.

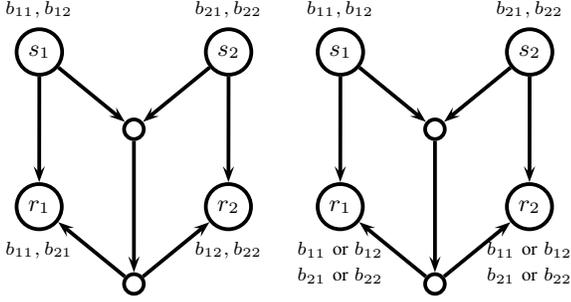

(a) Requesting specific messages.  (b) Requesting one message from each type.

Fig. 1: Example where each source has messages of a given type.

*Problem formulation:* We consider a network represented as a directed graph $G = (V, E)$, a set of $m$ sources $\{s_1, s_2, \cdots, s_m\}$ and a set of $n$ receivers $\{r_1, r_2, \cdots, r_n\}$. Each source has $u$ messages of the *same* type, and different sources have *different* types of messages. We denote by $\mathcal{B}_i = \{b_{i1}, b_{i2}, \cdots, b_{iu}\}$ the set of type-$i$ messages from source $s_i$.

*Performance metrics:* Given a transmission scheme, we use the rate towards a receiver to measure the efficiency of the transmission scheme. In the conventional message-specific coding problem, each receiver $r_j$ requests specific messages, one from each type, denoted by a fixed element $(b_1, b_2, \cdots, b_m) \in \mathcal{B}_1 \times \cdots \times \mathcal{B}_m$. For $T$ transmissions, the messages requested by receiver $r_j$ are denoted by $(b_1^T, b_2^T, \cdots, b_m^T)$, where each element is a vector $b_i^T = (b_i(1), \cdots, b_i(t), \cdots, b_i(T))$.

We denote by $\mathcal{R}_j$ the set of messages that the receiver $r_j$ can decode after $T$ transmissions, and we say that the transmission rate towards $r_j$ is the number of requested messages that can be decoded by the receiver $r_j$ per transmission, i.e.,

$$R_j = \frac{1}{T} \sum_{t=1}^{T} \sum_{i=1}^{m} I_{\{b_i(t) \in \mathcal{R}_j\}}. \quad (1)$$

In the content-type formulation, each receiver $r_j$ requests to receive (any) one message from each type, and does not care which specific one. We denote the requested content-type messages by an arbitrary element $x_1, x_2, \cdots, x_m \in \mathcal{B}_1 \times \cdots \times \mathcal{B}_m$. For $T$ transmissions, the messages requested by $r_j$ are denoted by $(x_1^T, x_2^T, \cdots, x_m^T)$, where each element is a vector $x_i^T = (x_i(t), \cdots, x_i(t), \cdots, x_i(T))$. Similarly, the rate towards $r_j$ is defined as the number of requested messages that can be decoded by the receiver $r_j$ per transmission, i.e.,

$$R_j^c = \frac{1}{T} \sum_{t=1}^{T} \sum_{i=1}^{m} I_{\{\exists x_i(t) \in \mathcal{R}_j, \text{ such that } x_i(t) \in \mathcal{B}_i\}}. \quad (2)$$

We would like to study the rate averaged among all receivers for message-specific coding, denoted by $R$, and the that for content type coding, denoted by $R^c$. Clearly, for message specific coding, the rate depends upon the specific message requests. We denote by $R^w$ the worst-case rate (minimizing among all possible sets of requests), and by $R^a$ the average rate (averaged over all possible sets of requests). We define the worst case and average case gains, respectively, as:

$$G^w = \frac{R^c}{R^w}, \quad G^a = \frac{R^c}{R^a}. \quad (3)$$

### A. Combination-like network

We consider the combination network structure $B(m, k, u)$, where $m \leq k$. shown in Fig. 2. The network has four layers: the first layer is the $m$ sources and each source connects to every node in the second layer; the second layer has $k$ intermediate $A$ nodes and each $A$ node connects to a $B$ node in the third layer, which also contains $k$ intermediate $B$ nodes; the fourth layer contains $n = \binom{k}{m} u^m$ receivers, where we have $u^m$ receivers connected to each subset of size $m$ of the $B$ nodes.

*Theorem 1:* In a $B(m, k, u)$ network,

$$G^w = u, \quad \lim_{k \to \infty} G^a = u. \quad (4)$$

The proof of this theorem follows from Lemmas 1-3.

*Lemma 1:* In the network $B(m, k, u)$, content type coding achieves $R^c = m$.

*Proof*: Use network coding to multicast one specific message from each type to all receivers. ∎

*Lemma 2:* In the network $B(m, k, u)$, the worst case message-specific coding rate is $R^w = m/u$.

*Proof*: We construct the following receiver requirements: for every $u^m$ receivers that are connected to the same $B$ nodes, each request is an element of the set $\mathcal{B}_1 \times \cdots \times \mathcal{B}_m$, and no two requests are the same. Since all $mu$ messages are requested, we need to use the same set of $m$ $A$-$B$ edges, at least $u$ times. Using network coding we can ensure that at the end of $u$ transmissions, each receiver gets all $mu$ messages and thus also the $m$ messages required by her; thus the transmission rate is $m/u$. Note that receiving all $mu$ messages by each receiver is a worst case scenario in terms of rate. ∎

*Lemma 3:* In a network $B(m, k, u)$, the average rate of the message-specific coding problem is bounded by

$$R^a \leq \frac{m}{u} + \frac{m}{u^{\frac{m+1}{2}}} + \frac{m^2(1+\sqrt{\ln u})}{u^{\frac{m+1}{2}}} + \frac{(m+1)^2\sqrt{\ln u}}{\sqrt{\binom{k}{m}} u^{\frac{m+1}{2}}}. \quad (5)$$

*Proof*: We here provide a short outline and give the complete proof in the appendix. We consider $m$ out of the $k$ edges that connect $A$ to $B$ nodes, and the $u^m$ receivers that can be reached only through these edges. We call these $m$ edges and $u^m$ receivers a basic structure. We argue that, through each such structure, we need to send almost surely all messages to have a good average - with high probability all messages are required by less than

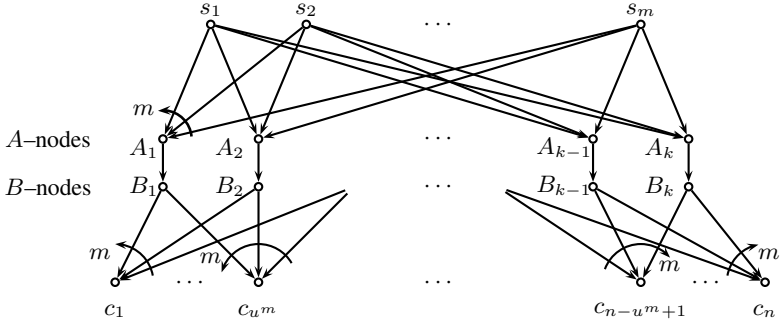

Fig. 2: Combination network structure.

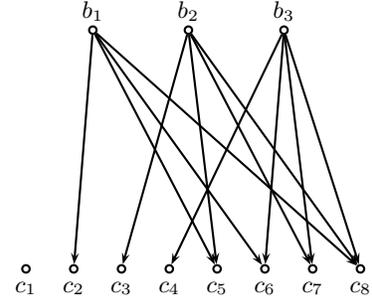

Fig. 3: A complete pliable coding instance.

$u^{m-1}(1-\delta_1)$ receivers. Thus the rate we can get through each basic structure converges to $m/u$. ∎

### III. CONTENT-TYPE CODING OVER ERASURE NETWORKS

We here make the case that, over erasure networks with feedback, we can realize benefits by allowing the random erasures to dictate the specific messages within a content type that the receivers get - essentially we shape what we serve to the random channel realizations.

We consider the following content-type coding setup. A server, $s$, has $k_1$ messages of content-type 1, $\mathcal{K}_1$, and $k_2$ messages of content-type 2, $\mathcal{K}_2$ (eg. an ad serving broadcasting station in a mall has $k_1$ sale coupons for clothes and $k_2$ sale coupons for gadgets). Receiver $c_1$ wants to receive all the $k_1$ sale coupons for clothes and some (a fraction $\alpha$, any $\alpha k_2$) of the sale coupons for gadgets; receiver $c_2$ wants the reverse, all the $k_2$ coupons for gadgets and some (any $\alpha k_1$) of the coupons for clothes. The server is connected to the two receivers through a broadcast erasure channel with state feedback; in particular, when the server broadcasts a message, each receiver $c_i$ receives it correctly with probability $1-\epsilon_i$, independently across time and across receivers. Each receiver causally acknowledges whether she received the message or not.

The corresponding message-specific scheme is as follows [6]. The server wants to send to $c_1$ all the messages in $\mathcal{K}_1$ and in a specific subset of $\mathcal{K}_2$, $\mathcal{K}_2^1 \subseteq \mathcal{K}_2$, of size $\alpha k_2$; and to $c_2$, all the messages in $\mathcal{K}_2$ and in a specific subset of $\mathcal{K}_1$, $\mathcal{K}_1^2 \subseteq \mathcal{K}_1$, of size $\alpha k_2$. We again have independent broadcast erasure channels with feedback.

*Rate region:* We say that rates $(r_1, r_2)$, with

$$r_1 = \frac{k_1 + \alpha k_2}{T}, \quad r_2 = \frac{k_2 + \alpha k_1}{T}, \quad (6)$$

are achievable, if with $T$ transmissions by $s$ both $c_1$ and $c_2$ receive all they require.

*Strategy for message-specific coding:* The work in [6] proposes the following achievability strategy and proves it is capacity achieving. Recall that we use the subscript to indicate the content type, and the superscript for the receiver.

• *Phase 1:* The source repeatedly transmits each of the messages in $\mathcal{K}_1 \setminus \mathcal{K}_1^2$ and $\mathcal{K}_2 \setminus \mathcal{K}_2^1$, until one (any one) of the two receivers acknowledges it has received it.

• *Phase 2:* The source transmits linear combinations of the messages in $\mathcal{K}_2^1$, $\mathcal{K}_1^2$, those in $\mathcal{K}_1 \setminus \mathcal{K}_1^2$ that were not received by $c_1$, and those in $\mathcal{K}_2 \setminus \mathcal{K}_2^1$ that were not received by $c_2$.

The intuition behind the algorithm is that, in Phase 1, each private message (that only one receiver requests) is either received by its legitimate receiver, or, it becomes side information for the other receiver. In Phase 2, each receiver either wants to receive each message in a linear combination or already has it as side information and can thus subtract it. The source creates linear combinations that are innovative (bring new information) to each receiver (eg., through uniform at random combining over a high enough field [6]). The strategy achieves the rate region:

$$\begin{cases} 0 \le r_1 \le \min\{1-\epsilon_1, \frac{1-\epsilon_2}{1-(1-\phi_2)(1-\alpha)}\} \\ 0 \le r_2 \le \min\{1-\epsilon_2, \frac{1-\epsilon_1}{1-(1-\phi_1)(1-\alpha)}\} \\ \frac{r_1}{1-\epsilon_1}\left(1-\frac{\alpha\phi_1}{1+\alpha}\right) + \frac{r_2}{1-\epsilon_{12}}\frac{1}{1+\alpha} \le 1 \\ \frac{r_1}{1-\epsilon_{12}}\frac{1}{1+\alpha} + \frac{r_2}{1-\epsilon_2}\left(1-\frac{\alpha\phi_2}{1+\alpha}\right) \le 1 \end{cases} \quad (7)$$

where $\epsilon_{12} = \epsilon_1 \epsilon_2$, and $\phi_i = (1-\epsilon_i)/(1-\epsilon_{12})$ for $i = 1, 2$.

*Strategy for content-type coding:* We propose the following strategy.

• *Phase 1:* For the messages in $\mathcal{K}_1$, denote by $\mathcal{K}_{1r}$ the messages not yet transmitted. Initially $\mathcal{K}_{1r} = \mathcal{K}_1$.

1) The server repeatedly broadcasts a message in $\mathcal{K}_{1r}$, until (at least) one of the receivers acknowledges it has successfully received it. The message is removed from $\mathcal{K}_{1r}$. If $c_1$ receives the messages, she puts it into a queue $Q_1^1$. If $c_2$ receives the message, she puts it into a queue $Q_1^2$.
2) The server continues with transmitting a next message in $\mathcal{K}_{1r}$, until either $\mathcal{K}_{1r}$ become empty, or

$$|\mathcal{K}_{1r}| + |Q_1^2| = \alpha k_1. \quad (8)$$

The server follows the same procedure for message set $\mathcal{K}_2$.

• *Phase 2:* The source transmits linear combinations of the messages in the three sets: $\mathcal{K}_{1r} \cup \mathcal{K}_{2r}$, $Q_1^2 \setminus Q_1^1$, and $Q_2^1 \setminus Q_2^2$ until both receivers are satisfied.

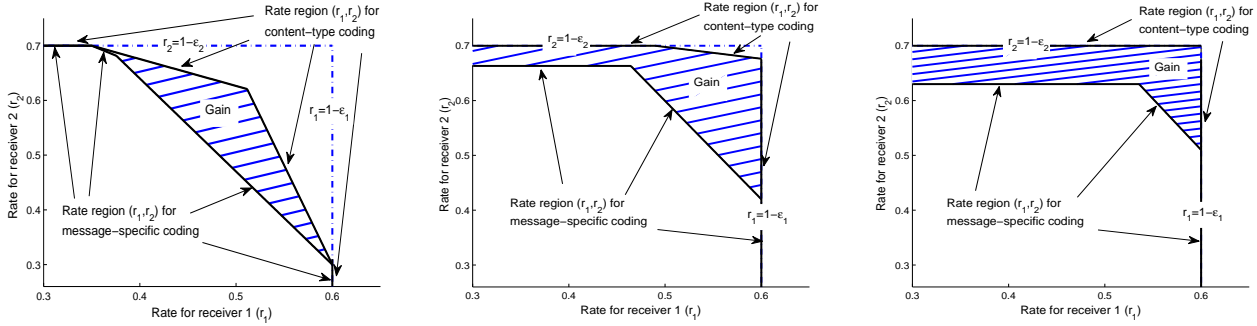

(a) Case 1: $\alpha < \min\{\phi_1, \phi_2\}; \alpha = 0.5$.  (b) Case 2: $\phi_1 < \alpha < \phi_2; \alpha = 0.7$.  (c) Case 3: $\alpha > \max\{\phi_1, \phi_2\}; \alpha = 0.85$.

Fig. 4: Comparison of rate region, as defined in (6), by message-specific and content-type coding, across three cases. The shaded regions show the gains of content-type over message-specific coding. The channel parameters are $\epsilon_1 = 0.4$ and $\epsilon_2 = 0.3$, which give $\phi_1 = 0.682$ and $\phi_2 = 0.795$.

The intuition behind the strategy is that, during Phase 1 we transmit messages from $\mathcal{K}_1$ until we either run out of messages, or both receivers want to receive the remaining $\mathcal{K}_{1r}$: $c_1$ because she wants all the messages in $\mathcal{K}_1$ and $c_2$ because, on top of the $Q_1^2$ she has already received, she also needs the $\mathcal{K}_{1r}$ to complete the fraction $\alpha k_1$. Note that originally, $|\mathcal{K}_{1r}| = k_1$ and $|Q_1^2| = 0$; at every step, the quantity in (8) either remains the same (if $c_2$ receives the message), or reduces by one (if she does not). Similarly for $\mathcal{K}_2$. In the second phase, the source again transmits linear combinations of messages that either a receiver wants, or already has and can subtract to solve for what she wants.

Using the above method, we can show the achievable rate of content-type broadcasting scheme in the following theorem.

*Theorem 2:* The rate region of the 1-2 content-type broadcasting communication with erasures is

$$\begin{cases} 0 \le r_1 \le \min\{1-\epsilon_1, \frac{1-\epsilon_2}{\alpha}\} \\ 0 \le r_2 \le \min\{1-\epsilon_2, \frac{1-\epsilon_1}{\alpha}\} \\ \frac{r_1}{1-\epsilon_1}[1 - \frac{\alpha(\phi_1-\alpha)^+}{1-\alpha^2}] + \frac{r_2}{1-\epsilon_1}\frac{(\phi_1-\alpha)^+}{1-\alpha^2} \le 1 \\ \frac{r_2}{1-\epsilon_2}[1 - \frac{\alpha(\phi_2-\alpha)^+}{1-\alpha^2}] + \frac{r_1}{1-\epsilon_2}\frac{(\phi_2-\alpha)^+}{1-\alpha^2} \le 1 \end{cases} \quad (9)$$

where $(x)^+ = \max\{x, 0\}$.

In fact, this achievable scheme achieves the capacity for 2-1 content-type broadcasting communication with erasures. The proof of achievability and converse of Theorem 2 is shown in Appendix B.

Fig. 4 compares the rate regions for the content-type and message coding. For content-type, we have three distinct cases, depending on the relative values of $\alpha$ and $\phi_i$. Note that $\phi_i$ expresses the fraction of messages that $c_i$ receives during Phase 1. Thus, if $\alpha < \min\{\phi_1, \phi_2\}$ (Fig. 4 (a)), $c_1$ and $c_2$ already receive $\alpha k_1$ and $\alpha k_2$ messages during Phase 1; essentially broadcasting content-type messages comes for "free", has not additional rate cost to providing $c_1$ with $\mathcal{K}_1$ and $c_2$ with $\mathcal{K}_2$. If $\min\{\phi_1, \phi_2\} < \alpha < \max\{\phi_1, \phi_2\}$, say for instance $\phi_1 < \alpha < \phi_2$ (Fig. 4 (b)), $c_2$ receives the content-type messages for free, but for $c_1$ we need additional transmissions in Phase 2. In $\alpha > \max\{\phi_1, \phi_2\}$ (Fig. 4 (c)), $c_1$ and $c_2$ require large percentages of messages from another type; interestingly, when we have $\max\{\phi_1, \phi_2\} < \alpha < \min\{\phi_1/\phi_2, \phi_2/\phi_1\}$, we can achieve the point $(1-\epsilon_1, 1-\epsilon_2)$, which implies that, all transmissions by $s$ are useful for both receivers. Message-specific coding in general does not achieve this point.

## IV. CONTENT-TYPE CODING IN THE INDEX CODING FRAMEWORK

Previous work has investigated a specific type of content-type coding within the framework of index coding, termed pliable index coding [7], [8]. In index coding we have a server with $m$ messages, and $n$ clients. Each client has as side-information a subset of messages, and requires one specific message. The goal is to minimize the number of broadcast transmissions so that each client receives the message she requested. In pliable index coding, the clients still have side information, but they no longer require a specific message: they are happy to receive any message they do not already have. The work in [7] [8] has shown that, although in the worst case, for index coding, we may require $O(n)$ transmissions, for pliable index coding we require at most $O(\log^2 n)$, i.e., we can be exponentially more efficient. This result was derived by using a bipartite representation of the pliable index coding problem and a randomized construction.

In this paper, apart from drawing attention to the fact that pliable index coding is a form of content-type coding, we also make two contributions: we derive an algebraic condition for clients to be satisfied in pliable index coding, and use this condition to prove a new lower bound: we show that there exist pliable index coding instances where $\Omega(\log n)$ transmissions are necessary.

*Bipartite graph representation:* We can represent a pliable index coding instance using an undirected bipartite graph, where on one side we have a vertex corresponding to each of the $m$ messages, say $b_1, \ldots, b_m$ and on the other side one vertex corresponding to each client, $c_1, \ldots, c_n$.

We connect with edges clients to the messages they *do not* have. For instance, in Fig. 3, client $c_5$ does not have (and would be happy to receive any of) $b_1$ and $b_2$.

### A. An algebraic criterion for pliable index coding

Assume that the source broadcasts $K$ linear combinations of the $m$ messages over a finite field $\mathbf{F}_q$; that is, the $k$-th transmission is $\alpha_{k,1} b_1 + \alpha_{k,2} b_2 + \cdots + \alpha_{k,m} b_m$, with $a_{i,j} \in \mathbf{F}_q$. We can collect the $K$ transmissions into a $K \times m$ coding matrix $\mathbf{A}$, where row $k$ contains the linear coefficients $(\alpha_{k,1}, \alpha_{k,2}, \cdots, \alpha_{k,m})$ used for the $k$-th transmission. We also denote by $\boldsymbol{\alpha}_i$ the $i$-th column vector of the matrix $\mathbf{A}$. Then each client receives $\mathbf{Ab} = \mathbf{c}$, where $\mathbf{b}$ is the $m \times 1$ vector that contains the messages and $\mathbf{c}$ is a constant vector, and needs using this and his side information to recover one message he does not have.

For client $c_j$, let us denote by $N[j]$ the set of indices of messages that $c_j$ does not have, and by $N_C[j]$ the set of indices of $c_j$'s side information. For example, $N[2] = \{1\}$ and $N_C[2] = \{2, 3\}$ for client $c_2$ in the example in Fig. 3. Clearly, client $c_j$ can use his side information to remove from the source transmissions the messages in $N_C[j]$; it thus can recover from matrix $\mathbf{A}$ the submatrix $\mathbf{A}_{N[j]}$ which only contains the columns of $\mathbf{A}$ with indices in $N[j]$. That is, if $\mathbf{b}_{N[j]}$ contains the messages he does not have, and $\mathbf{c}'$ is a constant vector, he needs to solve

$$\mathbf{A}_{N[j]} \mathbf{b}_{N[j]} = \mathbf{c}', \quad (10)$$

to retrieve any one message he does not have.

*Lemma 4:* In pliable index coding, client $c_j$ is satisfied by the coding matrix $\mathbf{A}$ if and only if there exists a column $\boldsymbol{\alpha}_i$, with

$$\boldsymbol{\alpha}_i \notin \mathrm{span}\{\mathbf{A}_{N[j] \setminus \{i\}}\}, \text{ for some } i \in N[j], \quad (11)$$

where $\mathrm{span}\{\mathbf{A}_{N[j] \setminus \{i\}}\} = \{\sum_{l \in N[j] \setminus \{i\}} \lambda_l \boldsymbol{\alpha}_l | \lambda_l \in \mathbf{F}_q, \boldsymbol{\alpha}_l \in \mathbf{F}_q^K\}$ is the linear space spanned by columns of $\mathbf{A}_{N[j]}$ other than $\boldsymbol{\alpha}_i$.

*Proof*: We are going to argue that, if such a column $\boldsymbol{\alpha}_i$ exists, then client $c_j$ can uniquely decode $b_i$ he does not have from his observations, and thus is satisfied. The condition $\boldsymbol{\alpha}_i \notin \mathrm{span}\{\mathbf{A}_{N[j] \setminus \{i\}}\}$ implies that any vector in the null-space $\mathcal{N}(\mathbf{A}_{N[j]})$ has a zero value in position $i$; indeed, since column $\boldsymbol{\alpha}_i$ is not in the span of the other columns, then for every vector $\boldsymbol{x} \in \mathcal{N}(\mathbf{A}_{N[j]})$, the only way we can have $\sum_{l \in N[j]} x_l \boldsymbol{\alpha}_l = 0$ is for $x_i = 0$. But the solution space of any linear equation can be expressed as a specific solution plus any vector in the nullspace; and thus from (10) we can get a unique solution for $b_i$ if and only if any vector $x$ in $\mathcal{N}(\mathbf{A}_{N[j]})$ has a zero value in the element corresponding to $i$, as is our case. We can retrieve $b_i$ by column and row transformations in (10). ∎

### B. A Lower Bound

*Theorem 3:* There exist pliable index coding instances that require $\Omega(\log(n))$ broadcast transmissions.

*Proof*: We constructively prove this theorem by providing a specific such instance, that we term the complete instance. In the complete instance, we have a client for every possible side information set corresponding to a client. In this case, the client vertex set $A$ corresponds to the power set $2^B$ of the message vertex set $B$, and we have $m = \log_2(n)$ (note that we can have an arbitrary number of additional messages and assume that all clients already have these). We define $W_d(A)$ to be the set of client vertices that have a degree $d$ $(d = 0, 1, \cdots, m)$, i.e., the vertices that need $d$ messages. An example of the complete instance with $m = 3$ is show in Fig. 3. Obviously, we can trivially satisfy all clients with $m = \log_2(n)$ transmissions, where each $b_i$ is sequentially transmitted once. We next argue that we cannot do better.

We will use induction to prove that the rank of the coding matrix $\mathbf{A}$ needs to be at least $m$ for the receivers to be satisfied according to Lemma 4. Let $J$ denote an index subset of the columns; in the complete model, Lemma 4 needs to hold for any subset $J$. For $|J| = 1$, i.e., to satisfy the clients who miss only one message, no column of the coding matrix $\mathbf{A}$ can be zero. Otherwise, if for example, column $i_1$ is zero, then the client who requests message $b_{i_1}$ cannot be satisfied. So $\mathrm{rank}(\boldsymbol{A}_J) = 1$ for $|J| = 1$. Similarly, for $|J| = 2$, any two columns of the coding matrix must be linearly independent. Otherwise, if for example, columns $i_1$ and $i_2$ are linearly dependent, then $\boldsymbol{\alpha}_{i_1} \in \mathrm{span}\{\boldsymbol{\alpha}_{i_2}\}$ and $\boldsymbol{\alpha}_{i_2} \in \mathrm{span}\{\boldsymbol{\alpha}_{i_1}\}$, and the clients who only miss messages $b_{i_1}$ and $b_{i_2}$ cannot be satisfied. So $\mathrm{rank}(\boldsymbol{A}_J) = 2$.

Suppose we have $\mathrm{rank}(\boldsymbol{A}_J) = l$ for $|J| = l$. For $|J| = l + 1$, we can see that if all clients who only miss $l + 1$ messages can be satisfied, then for some $i \in J$, we have $\boldsymbol{\alpha}_i \notin \mathrm{span}\{\boldsymbol{A}_{J \setminus \{i\}}\}$. Therefore, $\mathrm{rank}(\boldsymbol{A}_J) = \mathrm{rank}(\boldsymbol{\alpha}_i) + \mathrm{rank}(\boldsymbol{A}_{J \setminus \{i\}}) = 1 + l$. Therefore, to satisfy all the clients, the rank of the coding matrix $\boldsymbol{\alpha}$ is $m$, resulting in $K \geq m$, from which the result follows. ∎

## V. CONCLUSIONS AND SHORT DISCUSSION

This paper introduced a new problem formulation, that we termed content-type coding. Although there is significant work in content-distribution networks, the work still considers message-specific requests, where now multiple requests are interested in the same message. The research questions are focused mostly on where to store the content, how much to replicate, how much to compress, and what networks to use to upload and download content. There is also a rich literature in network coding and index coding, yet as far as we know, apart from the pliable index coding formulation, there has not been work that has looked at content-type coding.

We discussed in this paper three examples, where if we realize that we need to serve content-type rather than specific messages, we can have significant benefits. We believe that there are many more scenarios where we can realize benefits, as, downloading content-type rather than message-specific content, can help all aspects of content distribution networks, ranging from storage to coding to delivery. Even in the specific formulations we proposed in this paper, there are several followup directions and extensions, for instance looking at more than two users and multihop communication over erasure networks.

## APPENDIX A
### PROOF OF LEMMA 3

Let us denote by $\mathcal{E} = \{e_1, e_2, \cdots, e_k\}$ the set of $k$ edges connecting $A$ nodes and $B$ nodes. We refer to $m$ different edges in $\mathcal{E}$ and the $u^m$ receivers that are only connected to them, as a basic structure. From construction, there are $\binom{k}{m}$ such structures. For each structure, it is straightforward to see that:

- For any basic structure, the maximum transmission rate through these $m$ edges to the receivers in this structure is $m$ (since the min-cut is $m$).
- Denote by $s_{ij}$ the number of requirements of message $b_{ij}$ for receivers in a basic structure (i.e., the number of receivers requesting this message). Then the maximum transmission rate is $\sum_{l=1}^{m} \max_l s_{ij}/u^m$, where $\max_l s_{ij}$ represents the $l$-th maximum number among $\{s_{i,j} | 1 \leq i \leq m, 1 \leq j \leq u\}$. This means that the maximum transmission rate is achieved by transmitting the $m$ most popular messages through these $m$ edges.

Consider any given basic structure. We first observe that $E[s_{ij}] = u^{m-1}$. We define the event $E_{i,j}^{\delta_1} = \{s_{ij} > E[s_{ij}](1+\delta_1)\}$ as an abnormal event with respect to the message $b_{ij}$. We also define the event $E^{\delta_1} = \cup_{1 \leq i \leq m, 1 \leq j \leq u} E_{i,j}^{\delta_1}$ as an abnormal event for this basic structure. From the Chernoff bound, we have that

$$\Pr\{E_{i,j}^{\delta_1}\} = \Pr\{s_{ij} > E[s_{ij}](1+\delta_1)\} \leq e^{-\frac{u^{m-1}\delta_1^2}{3}}, \quad (12)$$

and

$$p_1 = \Pr\{E^{\delta_1}\} = um \Pr\{E_{i,j}^{\delta_1}\} \leq ume^{-\frac{u^{m-1}\delta_1^2}{3}}. \quad (13)$$

Here we denote by $p_1$ the probability that an abnormal event happens. When an abnormal event happens for this basic structure, the rate for this structure is (at most) $m$. If an abnormal event does not happen for this basic structure, the rate for this structure is (at most) $mu^{m-1}(1+\delta_1)/u^m = m(1+\delta_1)/u$.

Next, we consider the $v = \binom{k}{m}$ structures. Let us denote by $T_E$ the number of structures with an abnormal event happening. The expected value of $T_E$ is $vp_1$. The probability that $\{T_E > vp_1(1+\delta_2)\}$ happens, denoted by $p_2$, can be bounded using the Chernoff bound:

$$p_2 = \Pr\{T_E > vp_1(1+\delta_2)\} \leq e^{-\frac{vp_1\delta_2^2}{3}}, \quad (14)$$

Hence, if the event $\{T_E > vp_1(1+\delta_2)\}$ does not happen, the number of structures with an abnormal event is at most $vp_1(1+\delta_2)$. Therefore, the average rate can be bounded by

$$R^a \leq p_2 m + (1-p_2)[\frac{vp_1(1+\delta_2)m + (v-vp_1(1+\delta_2))\frac{m(1+\delta_1)}{u}}{v}]$$
$$\leq p_2 m + p_1(1+\delta_2)m + \frac{m(1+\delta_1)}{u}. \quad (15)$$

Let us set $\delta_1 = \frac{\sqrt{\frac{3}{2}(m+3)\ln u}}{u^{\frac{m-1}{2}}}$ and $\delta_2 = \frac{\sqrt{\frac{3}{2}(m+1)\ln u}}{\sqrt{vp_1}}$. Then we have

$$p_1 \leq mu^{-\frac{m+1}{2}},$$
$$p_2 \leq u^{-\frac{m+1}{2}}. \quad (16)$$

Plugging e.q. (16) into e.q. (15), we can have an upper-bound for the average rate:

$$R^a \leq p_2 m + p_1(1+\delta_2)m + \frac{m(1+\delta_1)}{u}$$
$$\leq \frac{m}{u} + \frac{m}{u^{\frac{m+1}{2}}} + \frac{m^2(1+\sqrt{\ln u})}{u^{\frac{m+1}{2}}} + \frac{(m+1)^2 \sqrt{\ln u}}{\sqrt{\binom{k}{m}}u^{\frac{m+1}{2}}}. \quad (17)$$

Setting $k = h_1 m$, and $m = h_2 u$, where $h_1$ and $h_2$ are constants, we have,

$$R^a \to \frac{m}{u}, \quad (18)$$

as $k \to \infty$. ∎

## APPENDIX B
### PROOF OF THEOREM 2

#### A. Achievability

To prove this theorem, we assume that $k_1$ and $k_2$ are large. Recall that we define $\epsilon_{12} = \epsilon_1 \epsilon_2$, and $\phi_i = (1-\epsilon_i)/(1-\epsilon_{12})$ for $i = 1, 2$. Let us denote by $k'_1$ and $k'_2$ the number of messages transmitted from sets $\mathcal{K}_1$ and $\mathcal{K}_2$ at the end of phase 1. Therefore, the average number of transmissions needed to complete phase 1 is:

$$N_1 = \frac{k'_1 + k'_2}{1 - \epsilon_{12}}. \quad (19)$$

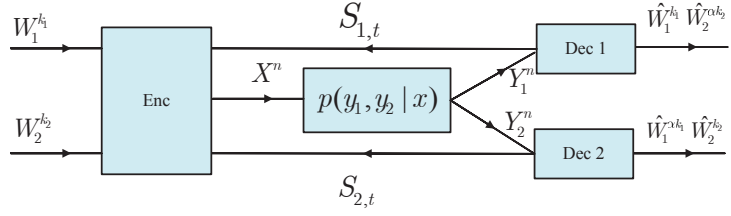

$$p(y_1, y_2 \mid x) = p(y_1 \mid x) p(y_2 \mid x) \quad p(y_i \mid x) = \begin{cases} x, & \text{wp } 1-\epsilon_i \\ e, & \text{wp } \epsilon_i \end{cases}$$

Fig. 5: Content type coding model for the broadcast erasure channel with feedback.

On average, the number of messages from set $\mathcal{K}_j$ ($j = 1, 2$) received by receiver $i$ ($i = 1, 2$) is:

$$M_j^i = k_j' \phi_i. \tag{20}$$

From the algorithm, we know that $k_1 - k_1' = (\alpha k_1 - M_1^2)^+$ and $k_2 - k_2' = (\alpha k_2 - M_2^1)^+$. Therefore, we have

$$k_i' = [1 - \frac{(\alpha - \phi_i)^+}{1 - \phi_i}] k_i. \tag{21}$$

In phase 2, the required number of messages for receiver $i$ ($i = 1, 2$) is then

$$k_r^i = (k_i' - M_i^i) + (k_1 - k_1') + (k_2 - k_2'), \tag{22}$$

where the first term is the number of erased messages from the set $\mathcal{K}_i$ that are received by another receiver, the second and third terms are the remaining messages to be transmitted.

For phase 2, the average number of transmissions needed is

$$N_2 = \max\{\frac{k_r^1}{1-\epsilon_1}, \frac{k_r^2}{1-\epsilon_2}\}. \tag{23}$$

Then, the rate region can be calculated as:

$$\{(r_1, r_2) : r_1 \geq 0, r_2 \geq 0, r_1 = \frac{k_1 + \alpha k_2}{T}, \\ r_2 = \frac{k_2 + \alpha k_1}{T}, N_1 + N_2 \leq T\}, \tag{24}$$

where $T$ is an auxiliary variable and can be cancelled out. Plugging (19) and (23) into (24), we get the theorem.

Note that for $\max\{\phi_1, \phi_2\} < \alpha < \min\{\phi_1/\phi_2, \phi_2/\phi_1\}$, the conditions are simplified as $r_1 \leq 1-\epsilon_1$ and $r_2 \leq 1-\epsilon_2$, implying that the maximum rates can be achieved.

*B. Converse*

To prove the converse of the theorem, we use an information theory method to show that this rate region is tight. We first depict the system model in Fig. 5. Recall that we denote by $s$ the transmitter, and $\mathcal{K}_1$ and $\mathcal{K}_2$ the sets of two types of messages. We denote by $k_1$ and $k_2$ the cardinalities of $\mathcal{K}_1$ and $\mathcal{K}_2$. Two receivers request information messages from the server $s$. Receiver 1 requests all information messages from $\mathcal{K}_1$ and $\alpha$ percent of the information messages from $\mathcal{K}_2$. In contrast, receiver 2 requests all information messages from $\mathcal{K}_2$ and $\alpha$ percent of the information messages from $\mathcal{K}_1$. Note that in our content type coding problem, receiver 1 is satisfied as long as she successfully receives $\alpha k_2$ messages from $\mathcal{K}_2$, disregarding of which ones, so is the same for receiver 2.

We consider broadcasting over an erasure channel with feedback. In this channel, when a message $x$ is transmitted, the receiver $i$ has a probability of $1 - \epsilon_i$ to receive the message correctly and has a probability of $\epsilon_i$ to receive nothing (i.e., the message is erased for receiver $i$). A message is received or erased independently for receivers 1 and 2. We aim to find the capacity region $(r_1, r_2)$ of for this broadcast channel. Without loss of generality, let us assume $|W_1| = |W_2| = |X| = |Y_1| - 1 = |Y_2| - 1 = 2$.

To prove the converse, we just need to show the first and the third equations in (9), and then according to the symmetry, we get the whole set of equations. For the first equation, it is equivalent to point-to-point communication, so we can directly get it. For the third equation, we consider two parts:

$$\frac{k_1}{1-\epsilon_1} + \frac{\alpha k_2}{1-\epsilon_1} \leq 1, \qquad \frac{k_1}{1-\epsilon_{12}} + \frac{k_2}{1-\epsilon_2} \leq 1,$$

First, we consider

$$\begin{aligned} n &\geq \sum_{t=1}^{n} H(X_t) \\ &\geq \sum_{t=1}^{n} H(X_t | Y_1^{t-1}, S^{t-1}) \\ &= \sum_{t=1}^{n} [H(X_t | Y_1^{t-1}, Y_2^{t-1}, S^{t-1}) \\ &\quad + I(X_t; Y_2^{t-1} | Y_1^{t-1}, S^{t-1}) \\ &= \sum_{t=1}^{n} [H(X_t | W_1^{k_1}, W_2^{k_2}, Y_1^{t-1}, Y_2^{t-1}, S^{t-1}) \\ &\quad + I(X_t; Y_2^{t-1} | Y_1^{t-1}, S^{t-1}) \\ &\quad + I(X_t; W_1^{k_1}, W_2^{k_2} | Y_1^{t-1}, Y_2^{t-1}, S^{t-1})] \end{aligned} \tag{25}$$

and the following two conditions:

$$\begin{aligned} (k_1 + k_2) - n\varepsilon_n &\leq I(W_1^{k_1}, W_2^{k_2}; Y_1^n, Y_2^n, S^n) \\ &= \sum_{t=1}^{n} I(Y_{1,t}, Y_{2,t}, S_t; W_1^{k_1}, W_2^{k_2} | Y_1^{t-1}, Y_2^{t-1}, S^{t-1}) \\ &= \sum_{t=1}^{n} I(Y_{1,t}, Y_{2,t}; W_1^{k_1}, W_2^{k_2} | Y_1^{t-1}, Y_2^{t-1}, S^{t-1}, S_t) \\ &= \sum_{t=1}^{n} I(X_t; W_1^{k_1}, W_2^{k_2} | Y_1^{t-1}, Y_2^{t-1}, S^{t-1}) \Pr(S_t \neq 0) \\ &= (1 - \epsilon_{12}) \sum_{t=1}^{n} I(X_t; W_1^{k_1}, W_2^{k_2} | Y_1^{t-1}, Y_2^{t-1}, S^{t-1}), \end{aligned} \tag{26}$$

where the above conditions hold due to the Fano's inequality and the incidence property of $S_t$, and

$$\begin{aligned}
\frac{k_1 - n\varepsilon_n}{1-\epsilon_1} &\leq I(W_1^{k_1}; X_t | Y_1^{t-1}, S^{t-1}) \\
&\leq \sum_{t=1}^n [I(X_t; W_1^{k_1} | Y_1^{t-1}, Y_2^{t-1}, S^{t-1}) \\
&\quad + I(X_t; Y_2^{t-1} | Y_1^{t-1}, S^{t-1})] \\
&\leq \frac{k_1}{1-\epsilon_{12}} + \sum_{t=1}^n [I(X_t; Y_2^{t-1} | Y_1^{t-1}, S^{t-1})],
\end{aligned} \quad (27)$$

where the last inequality follows from (using the same idea as (26))

$$\begin{aligned}
k_1 &\geq I(W_1^{k_1}; Y_1^n, Y_2^n, S^n) \\
&= (1-\epsilon_{12}) \sum_{t=1}^n I(X_t; W_1^{k_1} | Y_1^{t-1}, Y_2^{t-1}, S^{t-1}).
\end{aligned} \quad (28)$$

Plugging (26) and (27) into (25), we get the first part of the third equation in (9).

Similarly, we can get the second part of the third equation in (9) using

$$\begin{aligned}
(k_1 + \alpha k_2) - n\varepsilon_n &\leq I(W_1^{k_1}, W_2^{\alpha k_2}; Y_1^n, S^n) \\
&= (1-\epsilon_1) \sum_{t=1}^n I(X_t; W_1^{k_1}, W_2^{\alpha k_2} | Y_1^{t-1}, S^{t-1}) \\
&\leq (1-\epsilon_1) \sum_{t=1}^n H(X_t) \\
&\leq (1-\epsilon_1) n.
\end{aligned} \quad (29)$$